\documentclass[prd,twocolumn,amsmath,amssymb,nofootinbib,floatfix,superscriptaddress]{revtex4-1}

\usepackage{graphicx,bm}
\usepackage[normalem]{ulem} 
\usepackage{xcolor}
\newcommand{\mbf}[1]{\ensuremath{\boldsymbol{#1}}}

\begin{document}

\title{Flux tubes and string breaking in three dimensional SU(2) Yang-Mills
theory}

\author{Claudio Bonati} 
\email{claudio.bonati@unipi.it}
\affiliation{Dipartimento di Fisica dell'Universit\`a di Pisa 
       and INFN - Sezione di Pisa \\ Largo Pontecorvo 3, I-56127 Pisa, Italy}

\author{Silvia Morlacchi} 
\email{silviamorlacchi@gmail.com}
\affiliation{Dipartimento di Fisica dell'Universit\`a di Pisa 
       and INFN - Sezione di Pisa \\ Largo Pontecorvo 3, I-56127 Pisa, Italy}

\date{\today}

\begin{abstract}
We consider the three dimensional SU(2) Yang-Mills theory with adjoint static
color sources, studying by lattice simulations how the shape of the flux tube
changes when increasing the distance between them. The disappearance of the
flux tube at string breaking is quite abrupt, but precursors of this phenomenon
are present already when the separation between the sources is smaller than its
critical value, a fact that influences also some details of the static
potential.
\end{abstract}

\maketitle

\section{Introduction}\label{sec:intro}

Color confinement is one of the main nonperturbative features of nonabelian
gauge theories. A first principle proof of this phenomenon is still lacking and
constitutes part of the first Millennium Problem of the Clay Mathematical
Institute \cite{clay}. However a huge amount of information about color
confinement, both qualitative and quantitative, has been obtained by numerical
simulations of lattice discretized gauge theories.
 
Starting from the seminal work in Ref.~\cite{Fukugita:1983du}, the study of
flux tubes between static color charges gained a prominent role in the
investigation of color confinement \cite{DiGiacomo:1989yp, DiGiacomo:1990hc,
Bali:1994de, Haymaker:1994fm, Cea:1995zt}. The flux tube between two sources
has been investigated, for example, to test the predictions of effective string
theory \cite{Allais:2008bk, Gliozzi:2010zv, Gliozzi:2010jh, Cardoso:2013lla,
Amado:2013rja, Caselle:2016mqu} and the dual superconductor picture of color
confinement \cite{Cardaci:2010tb, Cea:2012qw, Cea:2014uja, Cea:2015wjd,
Baker:2018mhw, Battelli:2019lkz}, but also the case of more than two charges
has been studied \cite{Okiharu:2003vt, Bissey:2006bz, Bicudo:2011hk,
Bakry:2014gea}.
 
So far the vast majority of flux tube investigations concentrated on the
Yang-Mills pure glue case, with sources transforming in the fundamental
representation of the gauge group. In this setup a nonvanishing asymptotic
string tension is present\footnote{This is what happens for most of the gauge
groups used in the literature and, in particular, for the SU(N) gauge groups.
However gauge groups exist for which gluons do screen fundamental charges, and
when this happens the asymptotic string tension vanishes, see e.g.
\cite{Wellegehausen:2010ai} for the case of $G_2$.}, and the static potential
rises indefinitely with the distance between the sources; this signal that the
flux tube always connects them, independently of their distance. 

Only recently investigations carried out in full QCD with physical quark masses
appeared \cite{Cea:2017ocq, Bonati:2018uwh}, however the computational burden
of simulations with dynamical light flavours makes impossible to obtain in this
case results as accurate as those achieved for Yang-Mills theories.  In
particular, the QCD results obtained so far do not indicate any significant
qualitative difference with respect to the pure glue case.

Such a qualitative difference is however to be expected, since the asymptotic
string tension vanishes in theories with dynamical matter fields in the
fundamental representation of the gauge group. A striking consequence of this
fact is the peculiar behavior of the static potential: for small distances
between the sources the potential looks like the one of the pure glue case, but
when the separation increases beyond a critical value $R_c$ (the string
breaking length) the potential flattens, and does not grow anymore linearly
with the distance between the charges \cite{Wellegehausen:2010ai,
Philipsen:1998de, Detar:1998qa, Bernard:2001tz, Gliozzi:2004cs, Bali:2005fu,
Bulava:2019iut}.

It is natural to expect the flux tube to disappear, or at least to be strongly
suppressed, when the distance between the sources approaches $R_c$, but there
are several ways in which this could happen: the flux tube could for example
behave as in the pure glue case for small distances and then disappear abruptly
at $R_c$, or it could start to delocalize already when the sources are close to
each other. Which of these possibilities is the correct one can only be
established by numerical simulations, however to perform such a study in QCD
would be very demanding from the computational
point of view. We can nevertheless hope to gain at least some insight on what
happens in QCD by studying simplified models displaying string breaking. 

In this work we use for this purpose the three dimensional SU(2) Yang-Mills
theory with static sources transforming in the adjoint representation. It is
indeed simple to show that an adjoint charge can be screened by gluons, and
this model has been already used in the past to numerically investigate string
breaking and string decay in the static potential \cite{Stephenson:1999kh,
Philipsen:1999wf, Kratochvila:2003zj, Pepe:2009in} (see also \cite{Kallio:2000jc}
for the four dimensional case and e.g.  \cite{Agarwal:2007ns} for a non-lattice
approach). Our principal aim is the study of the flux tube behavior as a
function of the distance between the adjoint sources, and in particular for
distances close to the critical value $R_c$.  However, to better appreciate the
similarities and differences with respect to the case without string breaking,
we will also perform a precision study of the static potential in the unbroken
string phase. 

The paper is organized as follows: in Sec.~\ref{sec:setup} we summarize the
numerical setup adopted and we describe the observables used to study the flux
tube. Numerical results are reported in Sec.~\ref{sec:tube} and \ref{sec:pot}
for the flux tube and the static potential respectively. Finally in
Sec.~\ref{sec:concl} we summarize the results obtained and we draw our
conclusions.

\section{Numerical setup}\label{sec:setup}

As anticipated in the introduction, in this work we use the three-dimensional
SU(2) Yang-Mills theory with static sources in the adjoint representation as a
testbed to investigate the behavior of flux tubes close to string breaking.
The usual Wilson discretization \cite{Wilson:1974sk} is adopted, which for the
case of the gauge group SU(2) can be written in the form
\begin{equation}\label{eq:action}
S=\sum_{\mbf{x},\, \mu>\nu} \beta\left(1-\frac{1}{2}\Pi_{\mu\nu}(\mbf{x})\right)\ .
\end{equation}
In this expression $\mbf{x}$ is a point of a three dimensional isotropic
lattice with periodic boundary conditions, $\mu,\nu\in \{0, 1, 2\}$ denote
two lattice directions, and
\begin{equation}\label{eq:plaq}
\Pi_{\mu\nu}(\mbf{x})=\mathrm{Tr}\,\big[U_{\mu}(\mbf{x})U_{\nu}(\mbf{x}+\mbf{\hat{\mu}})
U^{\dag}_{\mu}(\mbf{x}+\mbf{\hat{\nu}})U^{\dag}_{\nu}(\mbf{x})\big]
\end{equation}
is the trace of the product of the link variables around the plaquette in
position $\mbf{x}$ laying in the plane $(\mu, \nu)$. The update is performed by
using standard heatbath \cite{Creutz:1980zw, Kennedy:1985nu} and microcanonical
\cite{Creutz:1987xi} moves, in the ratio of $1$ to $5$.

On the contrary of what happens in four-dimensional gauge theories, the gauge
coupling is not dimensionless in three space-time dimensions, and the bare
continuum coupling $g$ is related to the $\beta$ value entering
Eq.~\eqref{eq:action} by the relation $a\beta=4/g^2$, where $a$ denotes the
lattice spacing. As a consequence there is no dimensional transmutation in the
three-dimensional case, and dimensionless physical observables can be expanded
in inverse powers of $\beta$ in the weak coupling limit. In particular we will
sometimes use the following approximate expression for the square root of the
string tension \cite{Teper:1998te} 
\begin{equation}\label{eq:sigmateper}
a\sqrt{\sigma}=\frac{1.324(12)}{\beta}+\frac{1.20(11)}{\beta^2}+\mathcal{O}(\beta^{-3})\ ,
\end{equation}
which is valid for $\beta\ge 4.5$.  

The free energy (or the potential energy, in the zero temperature limit) of two static
adjoint color sources separated by a distance $d$ can be computed by using 
\begin{equation}
F^{\mathrm{adj}}(d)=-\frac{1}{aN_t}\log\langle\mathrm{Tr}P^{\mathrm{adj}}(\mbf{0}) 
\mathrm{Tr}P^{\mathrm{adj}}(d\mbf{\hat{1}})\rangle\ ,
\end{equation}
where $N_t$ is the temporal extent of the lattice, lattice translation and rotation
invariances have been used and $P^{\mathrm{adj}}(\mbf{x})$ denotes the adjoint
Polyakov loop in position $\mbf{x}$. The trace of $P^{\mathrm{adj}}(\mbf{x})$ can be
immediately related to the trace of the Polyakov loop in the fundamental
representation
\begin{equation}
P^{\mathrm{fund}}(\mbf{x})=\prod_{k=0}^{N_t-1}U_{0}(\mbf{x}+k\mbf{\hat{0}})
\end{equation} 
(where periodic boundary conditions are implied and 0 denotes the temporal
direction) by the relation
\begin{equation}
\mathrm{Tr}P^{\mathrm{adj}}(\mbf{x})=|\mathrm{Tr}P^{\mathrm{fund}}(\mbf{x})|^2-1\ .
\end{equation}

To investigate the flux tube between two static adjoint charges separated by a
distance $d$ along $\mbf{\hat{1}}$ (this choice of the direction is
purely conventional and irrelevant for the final result) we use the observable
\begin{equation}\label{eq:rhoadj}
\begin{aligned}
\hspace{-0.3cm}\rho_{\mu\nu}^{\mathrm{adj}}(d, x_t)=
\frac{\langle\mathrm{Tr}P^{\mathrm{adj}}(\mbf{0})
\mathrm{Tr}P^{\mathrm{adj}}(d\mbf{\hat{1}})
\Pi_{\mu\nu}\rangle}{
\langle\mathrm{Tr}P^{\mathrm{adj}}(\mbf{0})
\mathrm{Tr}P^{\mathrm{adj}}(d\mbf{\hat{1}})\rangle} - 
\langle\Pi_{\mu\nu}\rangle\ ,
\end{aligned}
\end{equation}
where $\Pi_{\mu\nu}$ stands for 
\begin{equation}\label{eq:plaqforrho}
\Pi_{\mu\nu}(d\mbf{\hat{1}}/2+x_t\mbf{\hat{2}})\ ,
\end{equation}
i.e. for the plaquette oriented in the $(\mu,\nu)$ plane, positioned midway
between the static sources at a transverse distance $x_t$.  On the lattice the
$d/2$ entering Eq.~\eqref{eq:plaqforrho} has obviously to be interpreted as the
integer division $\lfloor d/2\rfloor$. 
This ``midpoint'' flux tube is the one that has been most investigated in the
literature, mainly because in this way we minimize the effect of the static
charges. Of course it would be interesting to extend the study to have a
complete picture of the whole flux tube also closer to the static sources, but
this would require a decomposition of $\rho_{\mu\nu}^{\mathrm{adj}}$ in near
and far-field components (see \cite{Baker:2018mhw}).

$F^{\mathrm{adj}}(d)$ and $\rho_{\mu\nu}^{\mathrm{adj}}(d,x_t)$ are the natural
generalizations to the adjoint case of the usual expressions for fundamental
static sources, and it is simple to show that in the naive continuum limit
$\rho_{\mu\nu}^{\mathrm{adj}}$ reduces to the variation of $\langle
F_{\mu\nu}^2\rangle$ (no sum intended) induced by the presence of the adjoint
static sources. Moreover $\rho_{\mu\nu}^{\mathrm{adj}}$ is multiplicatively
renomalizable, and its renormalization constant is the same of $\Pi_{\mu\nu}$,
which also coincides with that of $\rho_{\mu\nu}^{\mathrm{fund}}$. In order to
avoid computing this renormalization constant we will use in the following the
ratio
\begin{equation}\label{eq:ratio}
R_{\mu\nu}(d, x_t)=\frac{\rho_{\mu\nu}^{\mathrm{adj}}(d, x_t)}{
\rho_{10}^{\mathrm{fund}}(d, 0)}\ ,
\end{equation}
which has a well defined continuum limit if numerator an denominator are
computed at the same lattice spacing.

To obtain accurate estimates of $F^{\mathrm{adj}}(d)$ and
$\rho^{\mathrm{adj}}_{\mu\nu}(d, x_t)$ we use both multihit
\cite{Parisi:1983hm} and multilevel \cite{Luscher:2001up} noise reduction
algorithms. The application of these algorithms is straightforward, once the
components of $P^{\mathrm{adj}}(\mbf{x})$ are explicitly written in term of
$P^{\mathrm{fund}}(\mbf{x})$ by using the relation
\begin{equation}
P^{\mathrm{adj}}_{ab}(\mbf{x})=\frac{1}{2}\mathrm{Tr}(\sigma_a P^{\mathrm{fund}}(\mbf{x}) 
\sigma_b [P^{\mathrm{fund}}(\mbf{x})]^{\dag})\ ,
\end{equation}
where $\sigma_a$ denotes a Pauli matrix. The optimal values for the number of
levels, the size of the slices and the number of updates to be used in the
multilevel algorithm has been determined by minimizing the fluctuations of
$\mathrm{Tr}\, P^{\mathrm{adj}}(\mbf{0}) \mathrm{Tr}\,
P^{\mathrm{adj}}(d\mbf{\hat{1}})$ at fixed simulation time. 

The optimal number of hits to be used in the multihit turned out to be quite
insensitive to the distance $d$ between the sources, while the optimal setup
for the multilevel algorithm typically consisted of a single level for small
distances between the sources, and of two levels for larger values of $d$. Let
us consider for example the case of the lattice $64^3$ at $\beta=11.3138$: the
setup adopted for $d=4a$ consisted of a single level algorithm with slices of
thickness $4a$ and 600 updates for slice, while for $d=15a$ we used two slices
of thickness $4a$ and $8a$, with 10000 and 10 updates for slice respectively.

In all the cases data corresponding to different values of $d$ and/or $x_t$
came from different simulations, and they are thus statistically independent of
each other. Statistical errors have been estimated by means of standard
blocking, jackknife and bootstrap procedures.

\section{Numerical results}\label{sec:res}

\subsection{Flux tube}\label{sec:tube}

In this section we report our results concerning the behavior of the flux tube
close to string breaking, obtained by studying the dependence of
$\rho_{\mu\nu}^{\mathrm{adj}}(d, x_t)$ (as a function of the transverse
distance $x_t$) on the separation $d$ between the adjoint static charges.  The
majority of our simulations have been performed on a $32^3$ lattice, but we
resorted also to different lattice sizes to investigate finite volume and
finite lattice spacing effects.

We mainly focus on the longitudinal component of the chromoelectric field
(corresponding to $\rho_{10}^{\mathrm{adj}}$ with the conventions of the
previous section), which turns out to be the dominant component of the flux
tube also in the adjoint case. However the study of the two other components of
the field strength is important to identify the disappearance of the flux tube:
since string breaking happens when the two charges are at a finite distance
from each other, we cannot expect the longitudinal chromoelectric field to
vanish at string breaking, because the near-field of the charges is always
present (see \cite{Baker:2018mhw}).  The natural expectation is that the
longitudinal component of the chromoelectric field became of the same size as
the other components at string breaking.

\begin{figure}[t] 
\centering 
\includegraphics[width=0.95\columnwidth, clip]{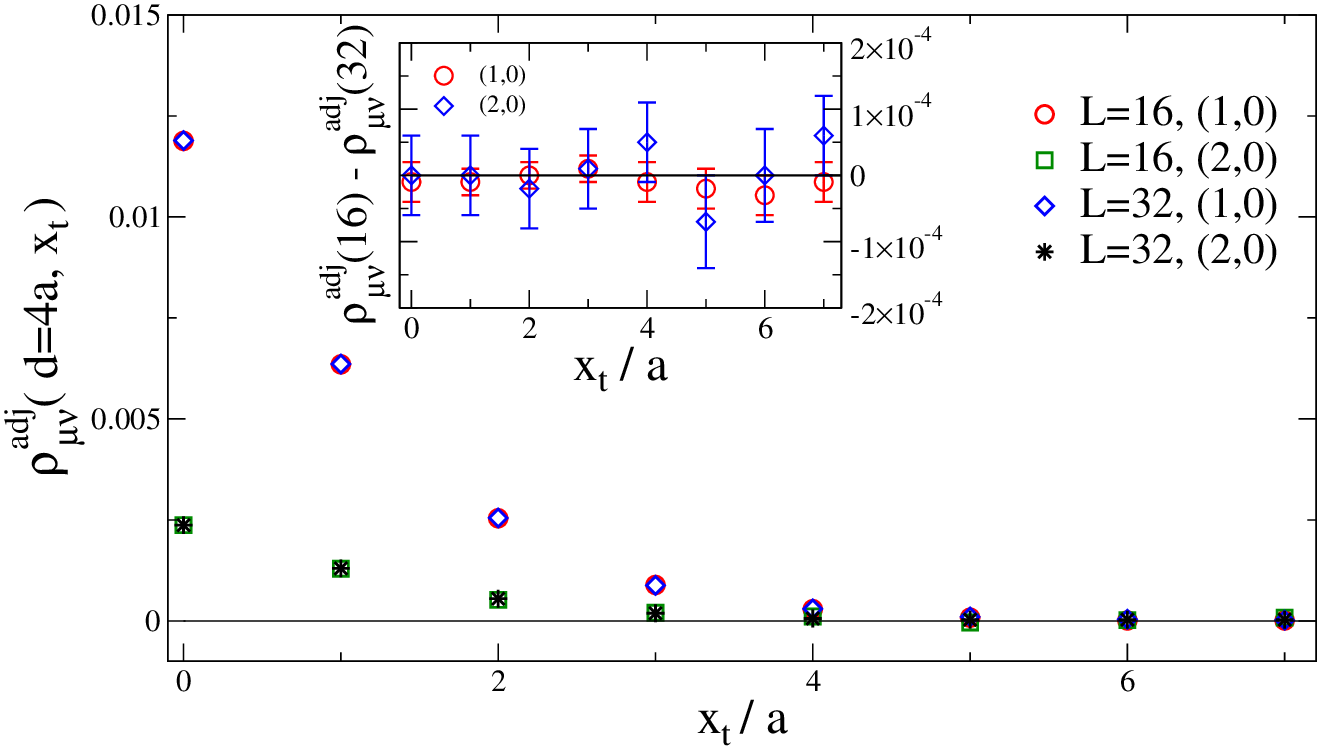}
\caption{Comparison of the estimates obtained for the quantity
$\rho_{\mu\nu}^{\mathrm{adj}}(d=4a,x_t)$ at $\beta=6.0$ by using different
lattice sizes ($L=16$ and $L=32$).  Results refer to the longitudinal $(1,0)$
and to the transverse $(2,0)$ components of the chromoelectric field.} 
\label{fig:finite_vol}
\end{figure}

As a first step we investigate which lattice sizes are needed in order not to
have significant finite volume effects. For this purpose we estimated
$\rho_{\mu\nu}^{\mathrm{adj}}(d,x_t)$ for $d=4a$ at coupling $\beta=6.0$, using
two different lattice sizes, i.e. $L=16$ and $L=32$. As can be seen from the
numerical results reported in Fig.~\ref{fig:finite_vol}, finite
size effects are well under control in this setup, and the longitudinal
component of the chromoelectric field is indeed the dominant component of the
flux tube. 

\begin{figure}[b] 
\centering 
\includegraphics[width=0.95\columnwidth, clip]{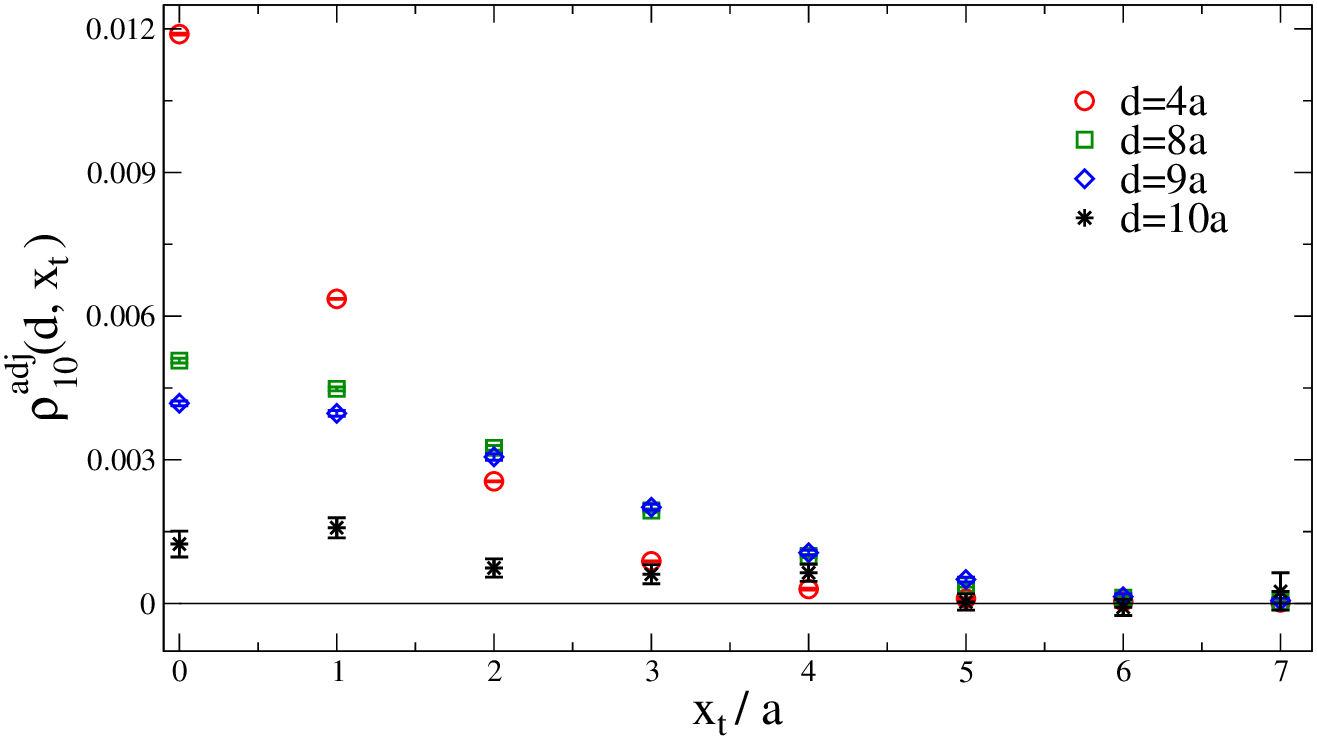}
\includegraphics[width=0.95\columnwidth, clip]{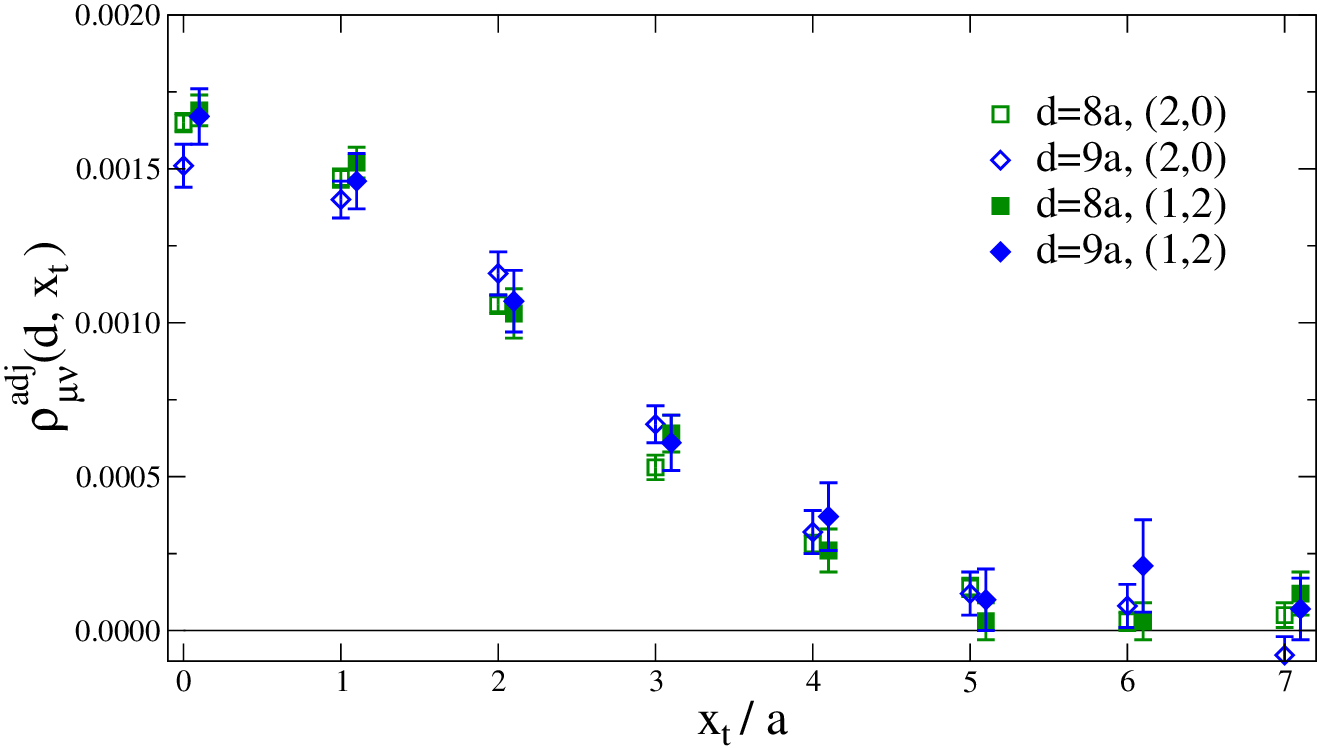}
\caption{Results for $\rho_{\mu\nu}^{\mathrm{adj}}(d,x_t)$ obtained on a $32^3$
lattice at coupling $\beta=6.0$.  In the upper panel the longitudinal component
is reported, while in the lower panel the transverse directions are shown (data
have been slightly shifted to improve the readability). Notice the different
scales on the vertical axis of the two panels.} 
\label{fig:tube_fix_scale}
\end{figure}

To study the dependence of the adjoint flux tube on the distance $d$ between
the static sources, we thus start by using a fixed scale approach on a $32^3$
lattice at $\beta=6.0$. For this value of the coupling 
the string breaking distance is approximately $R_c\approx 10a$ (see
\cite{Kratochvila:2003zj} and Sec.~\ref{sec:pot}), and results for
$\rho_{\mu\nu}^{\mathrm{adj}}(d,x_t)$ obtained in this setup are shown in
Fig.~\ref{fig:tube_fix_scale}, both for the longitudinal component
$\rho_{10}^{\mathrm{adj}}$ and for the transverse ones
$\rho_{20}^{\mathrm{adj}}$ and $\rho_{12}^{\mathrm{adj}}$. 

From data in Fig.~\ref{fig:tube_fix_scale} we can already draw several
interesting observations: first of all it is evident that the longitudinal
component of the adjoint flux tube decreases by increasing the distance between
the sources. While the huge decrease from $d=4a$ to $d=8a$ can be ascribed to
the closeness of the sources (and thus to the presence of the Coulomb component
at $d=4a$), the differences between $d=8a$ and $d=9a$ can not be interpreted in
this way. Indeed the transverse components of the field strength do not change
significantly, and the same happens for the flux tube in the fundamental
representation: for comparison $\rho_{10}^{\mathrm{fund}}(d, x_t=0)$ changes by
less than $4\%$ when going from $d=8a$ to $d=9a$, to be compared with the
$21\%$ change of $\rho_{10}^{\mathrm{adj}}(d, x_t=0)$. 

Another important thing to notice is that the longitudinal component of the
adjoint flux tube is about a factor three larger than the transverse components
at $d=8a$ and $9a$, however at string breaking (i.e. $d=10a$)
$\rho_{10}^{\mathrm{adj}}(d,x_t)$ suddenly drops and become compatible with the
transverse components. As previously discussed this is the smoking gun signal
of the flux tube disappearance, since for finite $R_c$ we can not expect the
longitudinal (or any other) component to vanish. A hint that at $d=10a$ the
physics of the system is changing comes also from the scaling of error-bars:
from Fig.~\ref{fig:tube_fix_scale} we see that $\rho_{10}^{\mathrm{adj}}(d,
x_t)$ data at $d=10a$ have errors which are approximately three times those at
$d=9a$, despite the fact that the statistics accumulated for $d=10a$ is about
six times larger than the one used for the other distances. A possible
interpretation of this fact is that for $d<10a$ the flux tube is present and
the main sources of statistical error in $\rho_{10}^{\mathrm{adj}}$ are the
fluctuations of Polyakov loops, which are however kept well under control by
using the multilevel algorithm. For $d=10a$ the string is broken and
fluctuations in the plane containing the plaquette increase (the ``broken
ends'' of the string moves freely), thus reducing the effectiveness of the
error reduction of our implementation of the multilevel algorithm.  

\begin{figure}[t] 
\centering 
\includegraphics[width=0.95\columnwidth, clip]{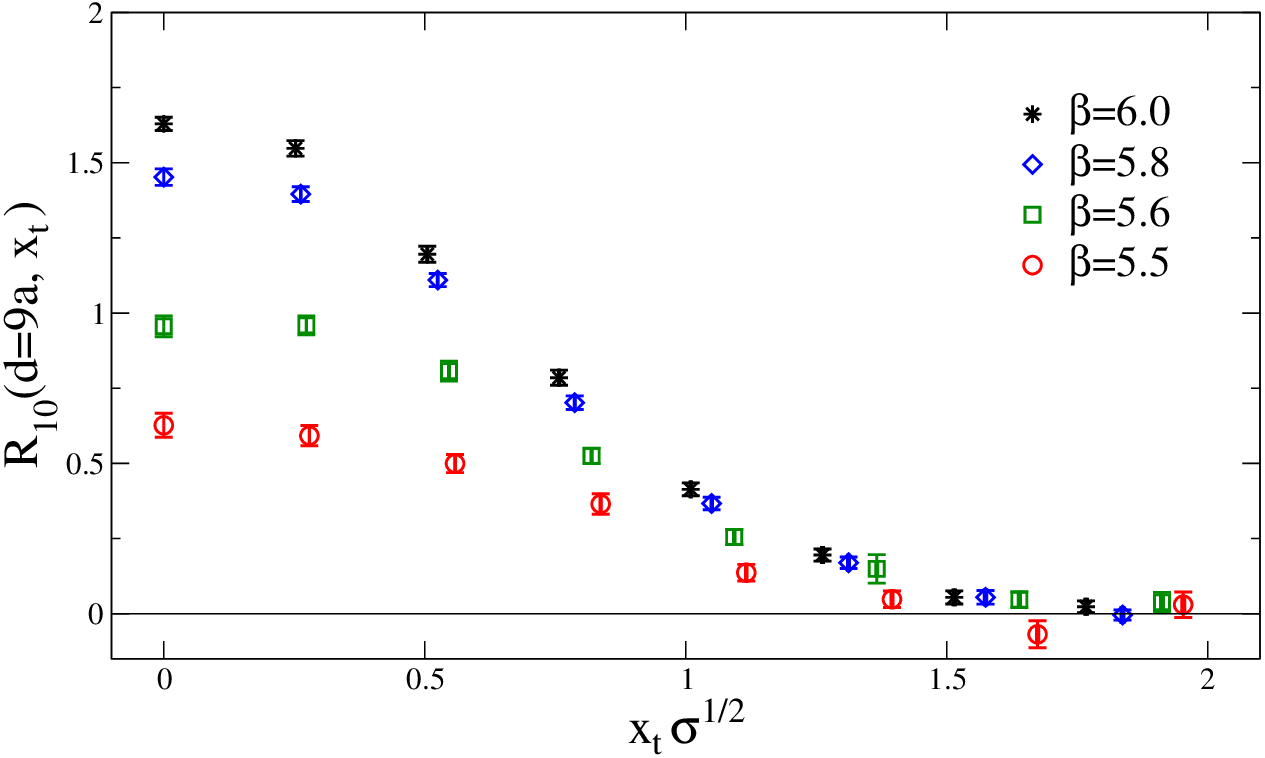}
\includegraphics[width=0.95\columnwidth, clip]{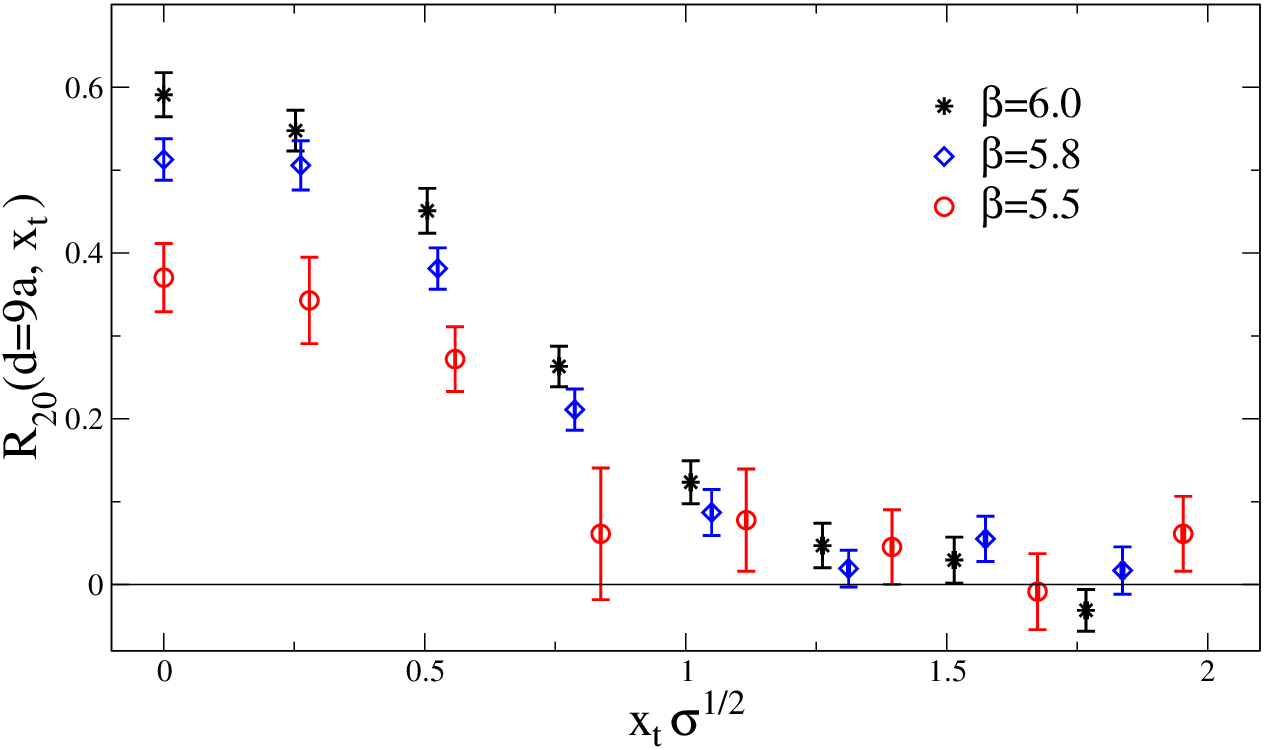}
\caption{Numerical results for the ratio $R_{\mu\nu}(d=9a, x_t)$
defined in Eq.~\eqref{eq:ratio} obtained using a $32^3$ lattice.} 
\label{fig:tube_varying_scale}
\end{figure}

\begin{table}[b]
\begin{tabular}{c|c|c}
$\beta$ & $a\sqrt{\sigma}$ & $a$ \\ \hline
5.5     & 0.2790(4)       & 0.12555(18) fm\\ \hline
6.0     & 0.2524(4)       & 0.11358(18) fm
\end{tabular}
\caption{String tension determined from the correlators of (fundamental)
Polyakov loops on a $32^3$ lattice. Conversion to physical units is performed
by using $1/\sqrt{\sigma}=0.45\,\mathrm{fm}$.}
\label{tab:sigma}
\end{table}

To have a finer control of the separation between the sources and better
resolve the distances close to string breaking, we now abandon the fixed scale
approach and change the distance between the sources by varying the lattice
spacing. More in detail we keep $d=9a$ on a $32^3$ lattice, and we increase the
lattice spacing by decreasing the value of the coupling constant $\beta$ in the
range $[5.5, 6.0]$. For $\beta=6.0$ and $5.5$ we explicitly computed the string
tension, obtaining the values reported in Tab.~\ref{tab:sigma}; these values
are consistent with those obtained by applying Eq.~\eqref{eq:sigmateper},
however their errors are significantly smaller than the ones we get from
Eq.~\eqref{eq:sigmateper}; the values of $a\sqrt{\sigma(\beta)}$ needed for
$5.5\le \beta\le 6.0$ are computed by using a linear interpolation of 
data in Tab.~\ref{tab:sigma}. 

To compare results obtained at different values of the lattice spacing we can
not use $\rho_{\mu\nu}^{\mathrm{adj}}$, due to the presence of the lattice
dependent renormalization, so we use the ratio $R_{\mu\nu}$ defined in
Eq.~\eqref{eq:ratio}. In Fig.~\ref{fig:tube_varying_scale} we present our
results for the longitudinal component $R_{10}(d=9a,x_t)$ at four different
values of the coupling $\beta$ in the range $[5.5, 6.0]$; some data for the
transverse component $(2,0)$ are also shown. 

As for the case of the fixed scale approach, we see from
Fig.~\ref{fig:tube_varying_scale} that the longitudinal component $R_{10}$ is
steeply decreasing when increasing the distance between the adjoint Polyakov
loops. In particular, its peak value at $x_t=0$ reduces approximately by a
factor of three when increasing the separation between the sources from
$\approx 1.02\,\mathrm{fm}$ (at $\beta=6.0$) to $\approx 1.13\,\mathrm{fm}$ (at
$\beta=5.5$). The transverse component $R_{20}$ also decreases when increasing
the lattice spacing, but in a less dramatic way than the longitudinal
component. 

\begin{figure}[t] 
\centering 
\includegraphics[width=0.95\columnwidth, clip]{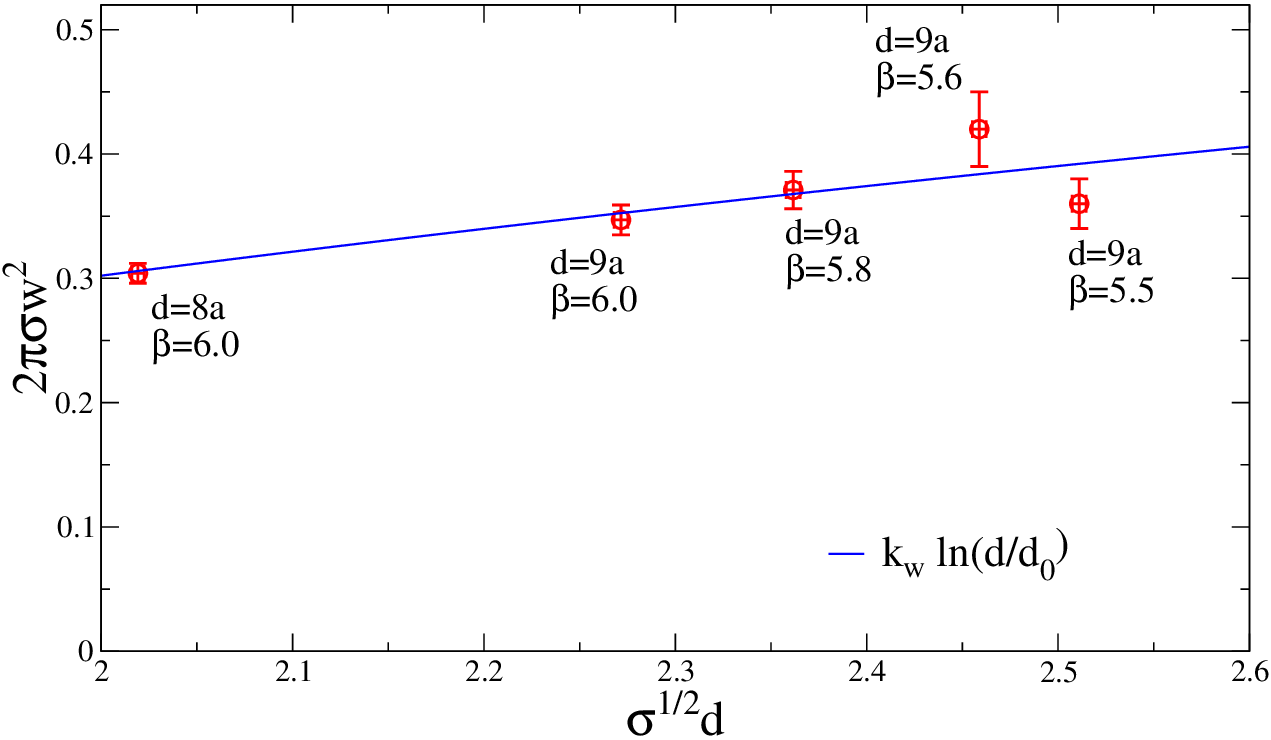}
\caption{Dependence of the width of the flux tube (as defined in
Eq.~\eqref{eq:wEST}) on the distance between the adjoint charges.}
\label{fig:tube_width}
\end{figure}

From Fig.~\ref{fig:tube_varying_scale} it is not completely clear if the
longitudinal flux tube just gets rescaled when approaching string breaking, or
it is also slightly distorted (i.e. the rescaling factor is different for
different values of $x_t$). To better investigate this point we tried
computing the flux tube width defined by
\begin{equation}
w^2(d)=\frac{\int_0^{\infty} x_t^2 \rho_{10}^{\mathrm{adj}}(d, x_t)\mathrm{d}x_t}{
\int_0^{\infty} \rho_{10}^{\mathrm{adj}}(d, x_t)\mathrm{d}x_t}\ .
\end{equation}
This quantity does not need any renormalization and it was computed by using a
spline interpolation of the data for $\rho_{10}(d, x_t)$.

Numerical results for $w^2(d)$ are shown in Fig.~\ref{fig:tube_width}, and a
slight increase of the flux tube width with the distance between the sources
seems to be present.  While there is no reason for Effective String Theory
(EST) to provide robust results for theories with string breaking, it is
nevertheless interesting to compare the observed behavior with the one
predicted by EST. In particular in Fig.~\ref{fig:tube_width} we also report the
result of a best fit of the form
\begin{equation}\label{eq:wEST}
2\pi\sigma w^2(d)=k_w\log(d/d_0)\ ,
\end{equation}
which for $k_w=1$ is the form expected on the basis of EST (see e.g.
\cite{Allais:2008bk}). The functional form in Eq.~\eqref{eq:wEST} well
describes numerical data for $w^2(d)$ but with $k_w=0.37(7)$, however the
dependence of $w^2(d)$ on the distance $d$ is mild enough that also a lineal
function correctly reproduces data. From this fact we can conclude that the flux
tube is not simply rescaled as $d$ approaches $R_c$, it 
gets slightly broader but the numerical accuracy is not enough to
reliably fix the functional form of $w^2(d)$.

\begin{figure}[t] 
\centering 
\includegraphics[width=0.95\columnwidth, clip]{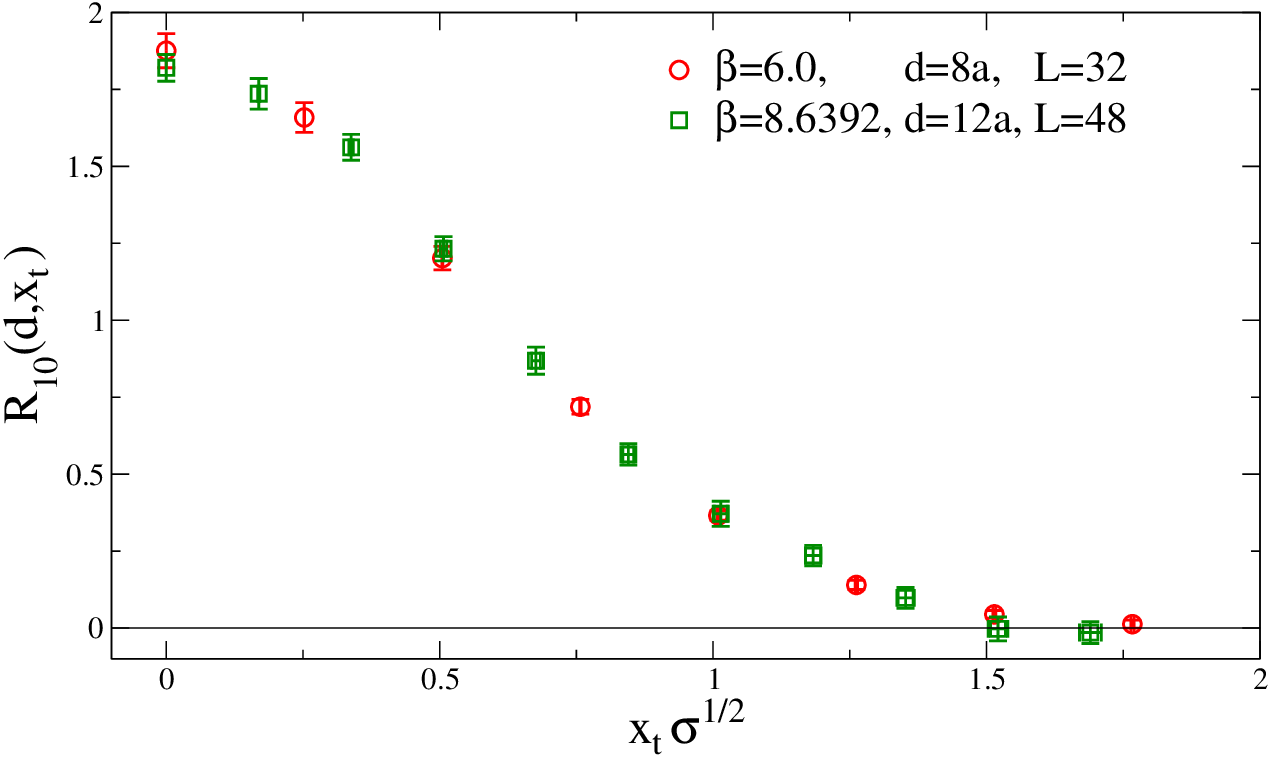}
\caption{Continuum scaling of $R_{10}(d, x_t)$ for $d\approx 0.91\,\mathrm{fm}$.}
\label{fig:cont_lim}
\end{figure}

To close this section we verify that lattice discretization artifacts do not
significantly affect the results presented so far. For this purpose we compare
data obtained by using two different lattice spacings, which have been
determined by using Eq.~\eqref{eq:sigmateper} to keep the value of $d$ constant
in physical units. We used a $32^3$ lattice at coupling $\beta=6.0$ and a
$48^3$ lattice with $\beta=8.6392$, in such a way that 
\begin{equation}
8a(\beta=6) \approx 0.91\,\mathrm{fm} \approx 12a(\beta=8.6392) \ . 
\end{equation}
The results obtained with this setup for $R_{10}(d, x_t)$ are presented in
Fig.~\ref{fig:cont_lim}, and it is clear that lattice artifacts are well under
control, being at most of the same size of statistical errors.

\begin{figure}[t] 
\centering 
\includegraphics[width=0.95\columnwidth, clip]{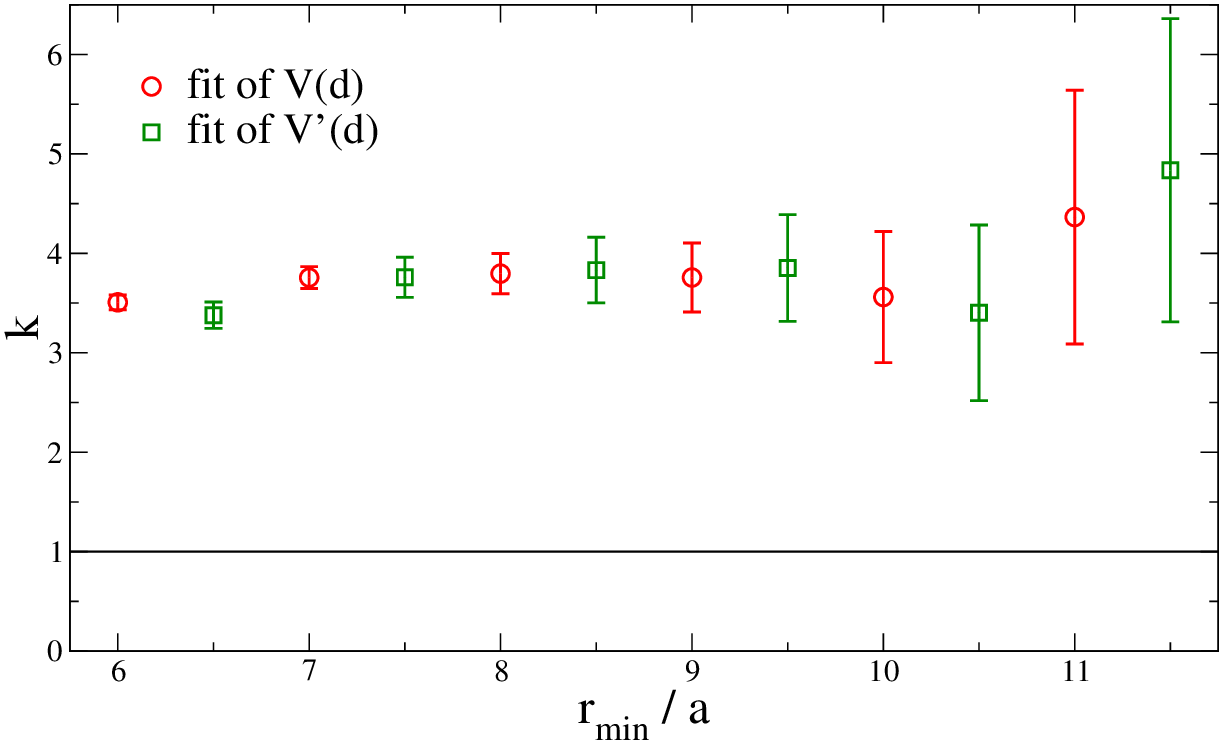}
\caption{Values of the coefficient $k$ defined in Eq.~\eqref{Vadj} obtained by
fitting on the interval $[r_{min},15a]$ data for $V^{\mathrm{adj}}$
(corresponding to integer values of $r_{min}/a$) or its first derivative
(corresponding to half-integer values of $r_{min}/a$). Data have been estimated
by using a $64^3$ lattice at $\beta=11.3138$.}
\label{fig:lusch_term}
\end{figure}

\subsection{Static potential}\label{sec:pot}

In this section we describe the results of our study of the static potential
between adjoint color charges, performed for distances between the sources
small enough to be in the unbroken string regime. The aim of this study is to
understand if the breaking of the string, associated to the dependence of the
flux tube on $d$ discussed in the previous section, has some
precursor in the behavior of the static potential.

One of the most typical properties of the static potential between fundamental
charges is the presence of the so called Luscher term \cite{Luscher:1980ac}.
This is just the first term of the EST expansion of the static potential in
powers of $\frac{1}{\sigma r^2}$ (see e.g. \cite{Brandt:2016xsp} for a recent
review), and it is characterized by the fact of having an universal
coefficient, which depends only on the space-time dimensionality but not on the
gauge group nor on other high-energy properties of the theory (as far as an
asymptotic string tension exists). In our three-dimensional setup the large
distance behavior of the fundamental static potential is thus
\begin{equation}
V^{\mathrm{fund}}(d)=\sigma d + \frac{\pi}{24 d} + \mathcal{O}(d^{-3})\ .
\end{equation}

Does something analogous to the Luscher term exist also for the static
potential $V^{\mathrm{adj}}(d)$ between adjoint sources?  While
$V^{\mathrm{adj}}$ has been previously investigated several times
\cite{Stephenson:1999kh, Philipsen:1999wf, Kratochvila:2003zj, Pepe:2009in}, to
the best of our knowledge an accurate investigation of the presence of the
Luscher term in $V^{\mathrm{adj}}$ has not been carried out so
far\footnote{This issue was mentioned in \cite{Kratochvila:2003zj} but the
Authors report that no stable fit parameter was found.}. We thus try to fit
data for $V^{\mathrm{adj}}$ according to the ansatz 
\begin{equation}\label{Vadj}
V^{\mathrm{adj}}(d) = \sigma d + k\frac{\pi}{24 d}\ ,
\end{equation}
where $k$ is a free parameter. Such an ansatz is reasonable only for $d<R_c$
however, just like in standard EST, values of $d$ which are too
small have to be excluded from the fit, since they are contaminated by the
Coulomb interaction between the sources (that in our case is logarithmic).

In Fig.~\ref{fig:lusch_term} we show our estimates for the parameter $k$
entering Eq.~\eqref{Vadj}, obtained by fitting data for $V^{\mathrm{adj}}(d)$
computed on a $64^3$ lattice at coupling $\beta=11.3138$.  According to
Eq.~\eqref{eq:sigmateper} the lattice spacing corresponding to this value of
the coupling is about half the one at $\beta=6.0$, so we expect $R_c\approx
20a$, and indeed up to $d=18a$ we found no signal of
string breaking. In Fig.~\ref{fig:lusch_term} we also report estimates obtained
by fitting the two-point finite difference approximation of the derivative of
$V^{\mathrm{adj}}$
\begin{equation}
\frac{\mathrm{d} V^{\mathrm{adj}}}{\mathrm{d}r}(r+a/2)\simeq
\frac{V^{\mathrm{adj}}(r+a)-V^{\mathrm{adj}}(r)}{a}\ 
\end{equation}
instead of the static potential itself, which give consistent results. From
Fig.~\ref{fig:lusch_term} we see that $k$ is definitely not
consistent with 1, and this fact can be interpreted as a signal for $d<R_c$
that the string will break by increasing the distance between the sources.

\begin{figure}[t] 
\centering 
\includegraphics[width=0.95\columnwidth, clip]{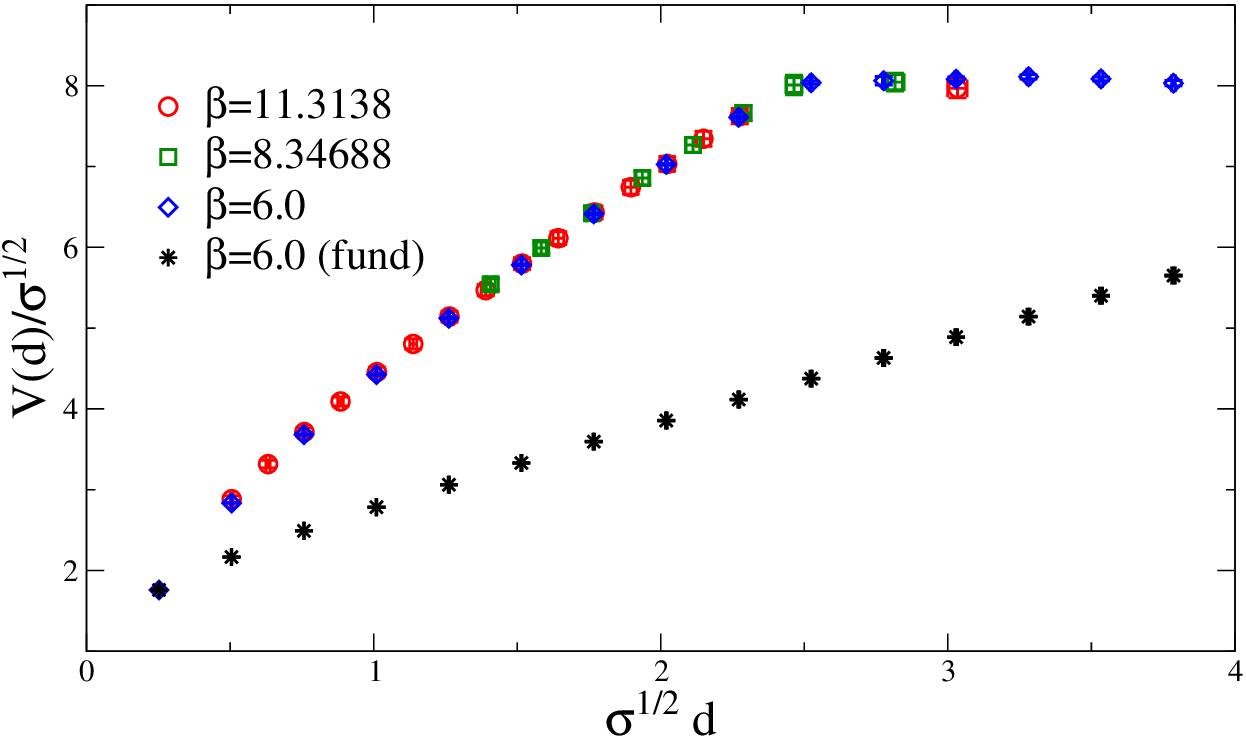}
\caption{Continuum scaling of the adjoint static potential: data have been
obtained on lattices with $L=32$ ($\beta=6.0$), $L=48$ ($\beta=8.34688$) and
$L=64$ ($\beta=11.3138$). The potential between two fundamental charges for
$\beta=6.0$ is also reported for comparison.}
\label{fig:adj_pot}
\end{figure}

Finally, in Fig.~\ref{fig:adj_pot} we show the continuum scaling of
$V^{\mathrm{adj}}$ for three different values of the lattice spacing (which
goes from $a\approx 0.11\,\mathrm{fm}$ at $\beta=6.0$ to $a\approx
0.057\,\mathrm{fm}$ at $\beta=11.3138$), with the static potential between
fundamental charges being also shown for comparison. Additive constants have
been fixed by imposing $V^{\mathrm{adj}}(2/\sqrt{\sigma})=7\sqrt{\sigma}$, and
an almost perfect scaling is observed, which implies also in this case the
absence of significant cut-off effects. 

\section{Conclusions}\label{sec:concl}

In this paper we have studied color flux tubes in a theory which displays string
breaking, and in particular their behavior when the separation between the
static sources approaches the string breaking distance $R_c$. For this
purpose we used as testbed the three-dimensional SU(2) Yang-Mills theory with
charges transforming in the adjoint representation of the gauge group.

We have shown that the adjoint flux tube, like the fundamental one, consists
mainly of the longitudinal chromoelectric field for distances $d$ between the
sources that are smaller than $R_c$. As the critical distance $R_c$ is
approached, the longitudinal chromoelectric field gets strongly suppressed,
becoming of the same size of the transverse fields at $R_c$. The disappearance
of the flux tube is quite abrupt, and the value of $R_{10}(d, x_t=0)$ (which is
related to square of longitudinal chromoelectric field inside the flux tube)
decreases approximately by a factor of $3$ when the relative difference between
$d$ and $R_c$ reduces below $10\%$. 

This rapid disappearance is the one that could have been naively guessed from
the behavior of the adjoint static potential $V^{\mathrm{adj}}(d)$, which
suddenly switches from an approximately linear grow to a constant plateau at
$d\simeq R_c$. We have however seen that precursors of string breaking are
present for $d$ smaller than $R_c$, which are basically related to the failure
of standard effective string theory.  The scaling of the square width $w^2(d)$
of the flux tube with the distance $d$ follows (at least within the present
accuracy) the expected logarithmic behaviour, but the value of the coefficient
differs from the universal effective string prediction. Similarly, an analogous
of the Luscher term is present also in $V^{\mathrm{adj}}(d)$, but again
numerical data are not compatible with the expected universal coefficient.

Future studies should be aimed at extending this analysis to other models, to
understand to which amount the phenomenology at string breaking observed in the
three-dimensional SU(2) Yang-Mills case is generic and, in particular, is
relevant for QCD. For the same reason it would be very interesting to
investigate if there is a relation between the values of the coefficients $k_w$
and $k$ in Eqs.~\eqref{eq:wEST},\eqref{Vadj} (or better, their deviations from
the EST predictions) and some nonuniversal property of the theory, like its
spectrum.

\emph{Acknowledgements} 
Numerical simulations have been performed on the CSN4 cluster of the Scientific
Computing Center at INFN-PISA.  It is a pleasure to thank Michele Caselle for
useful comments and discussions.

\end{document}